\documentclass[prl,reprint,superscriptaddress]{revtex4-1}

\usepackage{color}
\usepackage{graphicx}
\usepackage[colorlinks=true]{hyperref}
\usepackage{amsmath}
\usepackage{array}
\usepackage{subcaption}
\usepackage[justification=raggedright]{caption}
\begin{document}

\title{Dark Matter Results From 54-Ton-Day Exposure of PandaX-II Experiment}
\date{\today}
\author{Xiangyi Cui}
\affiliation{INPAC and School of Physics and Astronomy, Shanghai Jiao Tong University, Shanghai Laboratory for Particle Physics and Cosmology, Shanghai 200240, China}
\author{Abdusalam Abdukerim}
\affiliation{School of Physics and Technology, Xinjiang University, \"{U}r\"{u}mqi 830046, China}
\author{Wei Chen}
\author{Xun Chen}
\affiliation{INPAC and School of Physics and Astronomy, Shanghai Jiao Tong University, Shanghai Laboratory for Particle Physics and Cosmology, Shanghai 200240, China}
\author{Yunhua Chen}
\affiliation{Yalong River Hydropower Development Company, Ltd., 288 Shuanglin Road, Chengdu 610051, China}
\author{Binbin Dong}
\affiliation{INPAC and School of Physics and Astronomy, Shanghai Jiao Tong University, Shanghai Laboratory for Particle Physics and Cosmology, Shanghai 200240, China}
\author{Deqing Fang}
\affiliation{Shanghai Institute of Applied Physics, Chinese Academy of Sciences, Shanghai 201800, China}
\author{Changbo Fu}
\author{Karl Giboni}
\affiliation{INPAC and School of Physics and Astronomy, Shanghai Jiao Tong University, Shanghai Laboratory for Particle Physics and Cosmology, Shanghai 200240, China}
\author{Franco Giuliani}
\author{Linhui Gu}
\author{Yikun Gu}
\affiliation{INPAC and School of Physics and Astronomy, Shanghai Jiao Tong University, Shanghai Laboratory for Particle Physics and Cosmology, Shanghai 200240, China}

\author{Xuyuan Guo}
\affiliation{Yalong River Hydropower Development Company, Ltd., 288 Shuanglin Road, Chengdu 610051, China}

\author{Zhifan Guo}
\affiliation{School of Mechanical Engineering, Shanghai Jiao Tong University, Shanghai 200240, China}

\author{Ke Han}
\author{Changda He}
\author{Di Huang}
\affiliation{INPAC and School of Physics and Astronomy, Shanghai Jiao Tong University, Shanghai Laboratory for Particle Physics and Cosmology, Shanghai 200240, China}
\author{Shengming He}
\affiliation{Yalong River Hydropower Development Company, Ltd., 288 Shuanglin Road, Chengdu 610051, China}
\author{Xingtao Huang}
\affiliation{School of Physics and Key Laboratory of Particle Physics and Particle Irradiation (MOE), Shandong University, Jinan 250100, China}
\author{Zhou Huang}
\affiliation{INPAC and School of Physics and Astronomy, Shanghai Jiao Tong University, Shanghai Laboratory for Particle Physics and Cosmology, Shanghai 200240, China}
\author{Xiangdong Ji}
\email[Spokesperson: ]{xdji@sjtu.edu.cn}
\affiliation{Tsung-Dao Lee Institute, Shanghai 200240, China}
\affiliation{INPAC and School of Physics and Astronomy, Shanghai Jiao Tong University, Shanghai Laboratory for Particle Physics and Cosmology, Shanghai 200240, China}
\author{Yonglin Ju}
\affiliation{School of Mechanical Engineering, Shanghai Jiao Tong University, Shanghai 200240, China}

\author{Shaoli Li}
\author{Yao Li}
\author{Heng Lin}
\affiliation{INPAC and School of Physics and Astronomy, Shanghai Jiao Tong University, Shanghai Laboratory for Particle Physics and Cosmology, Shanghai 200240, China}

\author{Huaxuan Liu}
\affiliation{School of Mechanical Engineering, Shanghai Jiao Tong University, Shanghai 200240, China}

\author{Jianglai Liu}
\email[Corresponding author: ]{jianglai.liu@sjtu.edu.cn}
\affiliation{INPAC and School of Physics and Astronomy, Shanghai Jiao Tong University, Shanghai Laboratory for Particle Physics and Cosmology, Shanghai 200240, China}
\affiliation{Tsung-Dao Lee Institute, Shanghai 200240, China}

\author{Yugang Ma}
\affiliation{Shanghai Institute of Applied Physics, Chinese Academy of Sciences, Shanghai 201800, China}
\author{Yajun Mao}
\affiliation{School of Physics, Peking University, Beijing 100871, China}
\author{Kaixiang Ni}
\affiliation{INPAC and School of Physics and Astronomy, Shanghai Jiao Tong University, Shanghai Laboratory for Particle Physics and Cosmology, Shanghai 200240, China}
\author{Jinhua Ning}
\affiliation{Yalong River Hydropower Development Company, Ltd., 288 Shuanglin Road, Chengdu 610051, China}
\author{Xiangxiang Ren}
\affiliation{INPAC and School of Physics and Astronomy, Shanghai Jiao Tong University, Shanghai Laboratory for Particle Physics and Cosmology, Shanghai 200240, China}

\author{Fang Shi}
\affiliation{INPAC and School of Physics and Astronomy, Shanghai Jiao Tong University, Shanghai Laboratory for Particle Physics and Cosmology, Shanghai 200240, China}
\author{Andi Tan}
\email[Corresponding author: ]{anditan@umd.edu}
\affiliation{Department of Physics, University of Maryland, College Park, Maryland 20742, USA}

\author{Cheng Wang}
\affiliation{School of Mechanical Engineering, Shanghai Jiao Tong University, Shanghai 200240, China}

\author{Hongwei Wang}
\affiliation{Shanghai Institute of Applied Physics, Chinese Academy of Sciences, Shanghai 201800, China}

\author{Meng Wang}
\affiliation{School of Physics and Key Laboratory of Particle Physics and Particle Irradiation (MOE), Shandong University, Jinan 250100, China}

\author{Qiuhong Wang}
\email[Corresponding author: ]{wangqiuhong@sinap.ac.cn}
\affiliation{Shanghai Institute of Applied Physics, Chinese Academy of Sciences, Shanghai 201800, China}

\author{Siguang Wang}
\affiliation{School of Physics, Peking University, Beijing 100871, China}

\author{Xiuli Wang}
\affiliation{School of Mechanical Engineering, Shanghai Jiao Tong University, Shanghai 200240, China}

\author{Xuming Wang}
\affiliation{INPAC and School of Physics and Astronomy, Shanghai Jiao Tong University, Shanghai Laboratory for Particle Physics and Cosmology, Shanghai 200240, China}

\author{Qinyu Wu}
\affiliation{INPAC and School of Physics and Astronomy, Shanghai Jiao Tong University, Shanghai Laboratory for Particle Physics and Cosmology, Shanghai 200240, China}

\author{Shiyong Wu}
\affiliation{Yalong River Hydropower Development Company, Ltd., 288 Shuanglin Road, Chengdu 610051, China}

\author{Mengjiao Xiao}
\affiliation{Department of Physics, University of Maryland, College Park, Maryland 20742, USA}
\affiliation{Center of High Energy Physics, Peking University, Beijing 100871, China}

\author{Pengwei Xie}
\affiliation{INPAC and School of Physics and Astronomy, Shanghai Jiao Tong University, Shanghai Laboratory for Particle Physics and Cosmology, Shanghai 200240, China}

\author{Binbin Yan}
\affiliation{School of Physics and Key Laboratory of Particle Physics and Particle Irradiation (MOE), Shandong University, Jinan 250100, China}

\author{Yong Yang}
\affiliation{INPAC and School of Physics and Astronomy, Shanghai Jiao Tong University, Shanghai Laboratory for Particle Physics and Cosmology, Shanghai 200240, China}

\author{Jianfeng Yue}
\affiliation{Yalong River Hydropower Development Company, Ltd., 288 Shuanglin Road, Chengdu 610051, China}

\author{Dan Zhang}
\author{Hongguang Zhang}
\affiliation{INPAC and School of Physics and Astronomy, Shanghai Jiao Tong University, Shanghai Laboratory for Particle Physics and Cosmology, Shanghai 200240, China}

\author{Tao Zhang}
\author{Tianqi Zhang}
\author{Li Zhao}
\affiliation{INPAC and School of Physics and Astronomy, Shanghai Jiao Tong University, Shanghai Laboratory for Particle Physics and Cosmology, Shanghai 200240, China}

\author{Jifang Zhou}
\affiliation{Yalong River Hydropower Development Company, Ltd., 288 Shuanglin Road, Chengdu 610051, China}
\author{Ning Zhou}
\affiliation{INPAC and School of Physics and Astronomy, Shanghai Jiao Tong University, Shanghai Laboratory for Particle Physics and Cosmology, Shanghai 200240, China}
\author{Xiaopeng Zhou}
\affiliation{School of Physics, Peking University, Beijing 100871, China}

\collaboration{PandaX-II Collaboration}
\begin{abstract}
  We report a new search for weakly interacting massive particles
  (WIMPs) using the combined low background data sets acquired in 2016 and 2017
  from the PandaX-II experiment in China.  The latest data set
  contains a new exposure of 77.1 live day, with the background
  reduced to a level of 0.8$\times10^{-3}$ evt/kg/day, improved by a
  factor of 2.5 in comparison to the previous run in 2016. No excess
  events were found above the expected background. With a total
  exposure of 5.4$\times10^4$ kg day, the most stringent upper limit
  on spin-independent WIMP-nucleon cross section was set for a WIMP
  with mass larger than 100 GeV/$c^2$, with the lowest 90\% C.L. exclusion at
  8.6$\times10^{-47}$ cm$^2$ at 40 GeV/$c^2$.
\end{abstract}

\maketitle

Weakly interacting massive particles (WIMPs) are a leading class of
dark matter (DM) candidates~\cite{Bertone:2004pz} that have been
actively searched for in direct detection experiments using sensitive
detectors deployed in the deep underground laboratories around the
world. Liquid xenon detectors, such as those constructed and operated
by LUX, PandaX, and XENON collaborations, have been leading in
detection capability for heavy-mass WIMPs with masses larger than 10
GeV/$c^2$ all the way up to 100 TeV/$c^2$ or so
scale~\cite{Akerib:2016vxi, Tan:2016zwf, Aprile:2017iyp}, which is way
beyond the reach of the current generation of colliders. Their
detection sensitivities have reached the region predicted by popular
theory models (c.f. Ref.~\cite{Bagnaschi:2015eha}), which also leave
open a three-orders of magnitude discovery space for these exciting
experiments~\cite{Liu:2017drf}.

Located in the China Jinping Underground Laboratory
(CJPL)~\cite{Kang:2010zza}, the second phase of the PandaX project,
PandaX-II, is under operation since early 2016.  PandaX-II is a 580 kg
dual-phase xenon time-projection chamber (TPC), with a 60$\times$60~cm
cylindrical target viewed by 55 top and 55 bottom Hamamatsu R11410-20
3-inch photomultiplier tubes (PMTs), collecting both the prompt
scintillation photons ($S1$) in the liquid and the delayed
proportional scintillation photons ($S2$) in the gas. The first
low-background physics run (Run 9) collected a DM search data for 79.6
live days in 2016, and a world-leading result was published in
Ref.~\cite{Tan:2016zwf}, in combination with the data obtained during
the commissioning period~\cite{Tan:2016diz}.  In this paper, we report an updated
WIMP search using Run 9 and a new 77.1 live days data set taken in
2017 (Run 10) with significantly lower background level. This paper presents
one of the most sensitive WIMP searches using the largest direct
detection exposure to date.

Prior to Run 10, we made an extended calibration for the
electron-recoil (ER) events using CH$_3$T, a technique pioneered by
the LUX collaboration~\cite{Akerib:2015wdi}. Although the getter was
effective in removing the tritium afterwards, the tritium decay rate
plateaued at a rate of 2.0$\pm$0.4 $\mu$Bq/kg, which strongly
indicated tritium attachment on detector surfaces and their slow
emanation. In order to eliminate the tritium and to further reduce the
krypton background, the detector was emptied and reconditioned, and a
re-distillation of xenon was carried out on site. In Feb. 2017, the
detector was re-filled, and Run 10 dark matter search data set was
collected between April to July.


In Run 10, a number of detector PMTs ran at a lower gain than
previously, beyond which discharge signals would show up, and as a
positive consequence, the average dark rate per PMT was reduced from
1.9~kHz to 0.17~kHz.  To calibrate the gains of the PMTs, very low
intensity blue LED runs were taken with the digitizers in
full-recording mode (no baseline suppression) twice a week.
During the regular data taking, a value of 20 ADC counts relative to
the baseline, corresponding to an amplitude of about 0.4 single
photoelectron (SPE) in Run 9, and 0.6 SPE (due to lower gain) in Run
10, was set as the threshold below which the ``zero-length encoding''
(ZLE) firmware of the CAEN V1724 digitizers suppressed the data
recording~\cite{v1724_reference}.
The inefficiency of ZLE to SPE was studied channel-by-channel using
the LED calibration runs by comparing the detected photoelectron area
with and without the ZLE. The overall efficiency summing over all PMTs
for a detected $S1$-like signal of 3 PE (lower selection window for DM),
was 91\% and 78\%, respectively, during Run 9 and Run
10, indicating that lower gain led to lower ZLE efficiency (see Fig. 6
in Supplemental Material~\cite{sup_material}). This is one
significant source of inefficiency that must be taken into account in
signal and background modeling.

Out of 110 3-inch PMTs, one top and two bottom ones were kept off in
Run 10 (only one bottom PMT was inactive in Run 9).  In addition,
another peripheral top PMT was noisy, hence a high ZLE threshold was
set under the cost of its low efficiency for small pulses.
The cathode high voltage (HV) was also lowered to $-24$ kV ($-29$ kV
in Run 9) to avoid spurious discharges. The gate HV was maintained at
$-$4.95 kV, same as in Run 9. The maximum drift time for electron
changed from 350~$\mu$s in Run 9 to 360~$\mu$s in Run 10.  A FPGA-based
trigger system was implemented to replace the analog trigger, which
reduced the trigger threshold for $S2$ to about 50
PE~\cite{Wu:2017cjl}.
The electron lifetime in Run 10 was improved to an average of
850~$\mu$s, compared to 623~$\mu$s in Run 9.

\begin{figure}
  \centering
  \includegraphics[width=0.95\linewidth]{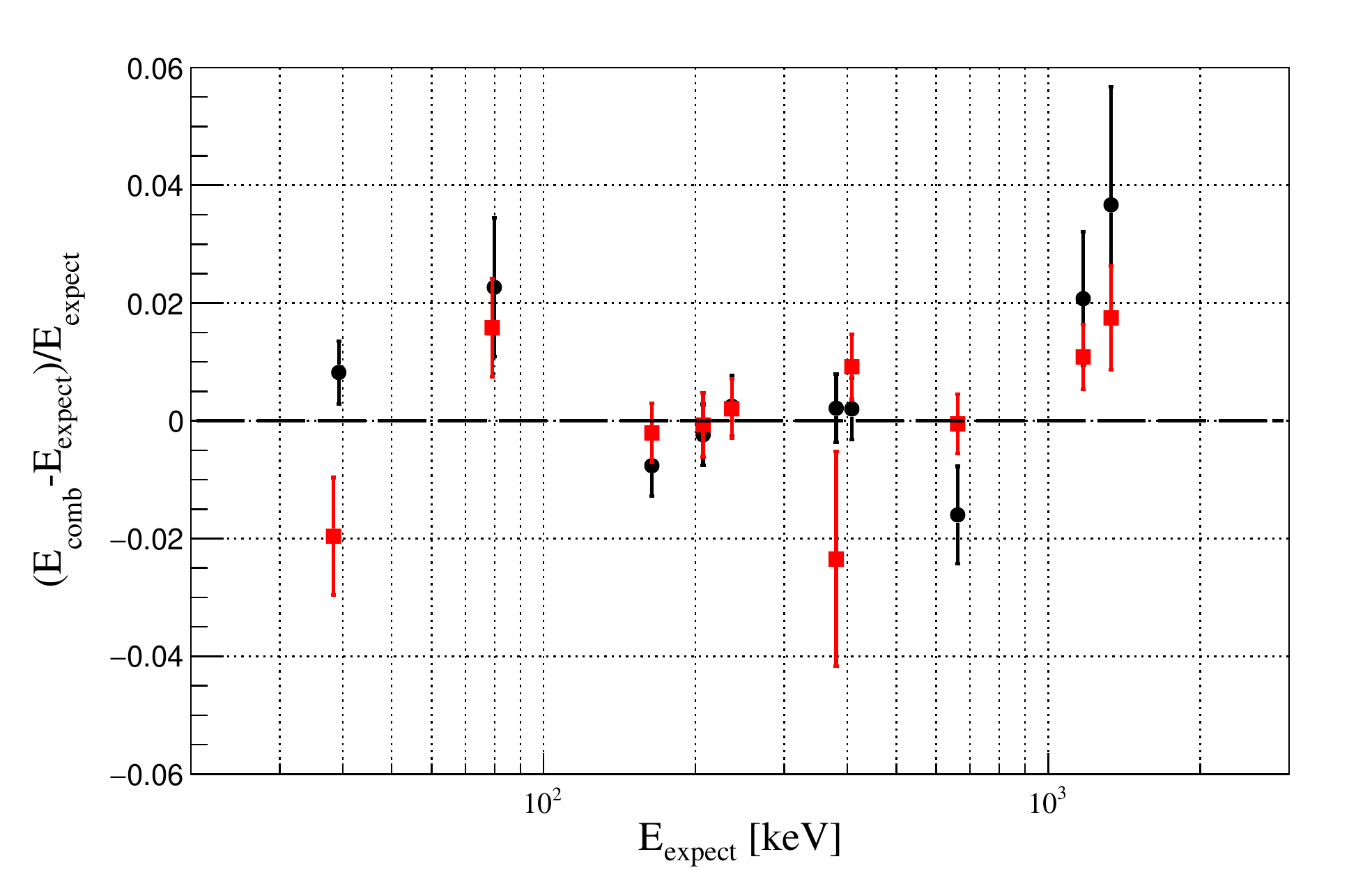}
  \caption{Fractional difference between the reconstructed energy
    $E_{\rm{comb}}$ and expected energy $E_{\rm expect}$ vs. $E_{\rm
      expect}$ for ER peaks in Run 9 (black circles) and Run 10 (red
    squares), using the best fit PDE and EEE values described in the
    text. Uncertainties include both statistical and systematic
    components.}
  \label{fig:Ecomb}
\end{figure}

Similar to Ref.~\cite{Tan:2016zwf}, corrections
to $S1$ and $S2$ were made using the position-dependence of $^{131\rm{m}}$Xe
de-excitation peak throughout the detector.  The detector responses to
high energy ER peaks, including 39.6 keV (n, $^{129}$Xe$^*$), 80.2 keV
(n, $^{131}$Xe$^*$), 164 keV ($^{131\rm{m}}$Xe), 236 keV
($^{129\rm{m}}$Xe), 408 keV ($^{127}$Xe), 662 keV ($^{137}$Cs), and
1173 keV ($^{60}$Co), were used to determine the overall photon and
electron detection efficiency. As in Ref.~\cite{Tan:2016zwf}, the energy is
reconstructed as
\begin{equation}
E_{\rm{comb}} = 0.0137\,{\rm{keV}}
\left(\frac{S1}{\rm{PDE}}+\frac{S2}{\rm{EEE}\times\rm{SEG}}\right),
\end{equation}
where PDE, EEE, and SEG are photon detection efficiency, electron
extraction efficiency, and single electron gain, respectively. The SEG
was determined using the charge distribution from the smallest $S2$ to
be 24.4$\pm$0.7 (Run 9) and 23.9$\pm$0.5 (Run 10) PE/$e^{-}$, with ZLE
efficiency taken into account.  For ER energy exceeding 200 keV, there
were non-negligible saturation effects from the digitizers and the
PMTs, for $S2$ signal in particular.  Like in the previous analysis,
we used $S2_b$ ($S2$ from the bottom array) to reconstruct high energy
events. However, in this analysis, $S2_b$ was corrected with a new $S2_b$
uniformity correction, different from the total $S2$ uniformity
correction appropriate for the lower energy events. The saturation of
$S2_b$ as a function of vertical position was also taken into account.
In addition, instead of making cuts on
$S1$ and $S2$ to select ER peaks, we took all data and performed a
parameter scan in PDE and EEE to fit $E_{\rm{comb}}$ peaks to the
expected energies. The resulting best fit $E_{\rm comb}$ agreed with
their expectation within 2\% for the entire energy range considered
(Fig.~\ref{fig:Ecomb}).
The updated PDE and EEE were 11.14$\pm$0.78\% and 54.5$\pm$2.7\% for
Run9 and 11.34$\pm$0.46\% and 57.7$\pm$1.9\% for Run 10. Minor ZLE effects
have been taken into account for the ER peaks in this
analysis. Consistent values were obtained using low energy tritium events. 

The neutron calibration data using AmBe source
(Ref.~\cite{Tan:2016zwf}) were re-analyzed with a significant
improvement in modeling.  Some neutrons multiple-scatter with single
energy deposition in the sensitive region and a partial energy
deposition in the below-cathode region, where the electrical field
direction is opposite to the drift field in the target volume so the
corresponding charge could not be detected. There are the so-called
``neutron-X'' events with abnormally suppressed $S2/S1$, which mix
with the pure single scatter nuclear-recoil (NR) events.  The vertical
distribution in the calibration data was fitted with that expected
from the Geant4-based~\cite{Agostinelli:2002hh,Allison:2006ve} Monte
Carlo (MC) simulation, from which the rates of the two components were
separated statistically.  Another custom software then simulated the
distribution of these events in $S1$ and $S2$.  This simulation began
with detector field configuration, and then proceeded the
photon/electron productions and fluctuations based on the NEST
framework~\cite{Lenardo:2014cva}. It also incorporated the measured
$S1$ and $S2$ spatial non-uniformity, channel-by-channel hit probability,
double PE emissions~\cite{Faham:2015kqa} (probability measured to be
0.22$\pm$0.02), ZLE effect, and the number-of-fired PMT $\ge 3$ cut. A
tuning on a single parameter, $\alpha=N_{ex}/N_i$, the ratio of the
initial excitation to ionization, was performed to the NEST model, to
match the 2D distribution in ($S1$, $S2$) between data and simulation
for $S1>5$ PE and $S2_{\rm raw}>200$ PE.  The NR detection efficiency
was then obtained by comparing the data and simulation.  In the
``plateau region'' for $S1>7$~PE and $S2_{\rm raw}>200$~PE, the
efficiency was anchored to 94\%, derived from number of events in the
central part of AmBe band before and after data quality cuts.  The
tuned NR response model and its detection efficiency was used to
produce the probability density functions (PDFs) for DM signals as
well as the neutron background.

\begin{figure}
  \centering
  \includegraphics[width=0.95\linewidth]{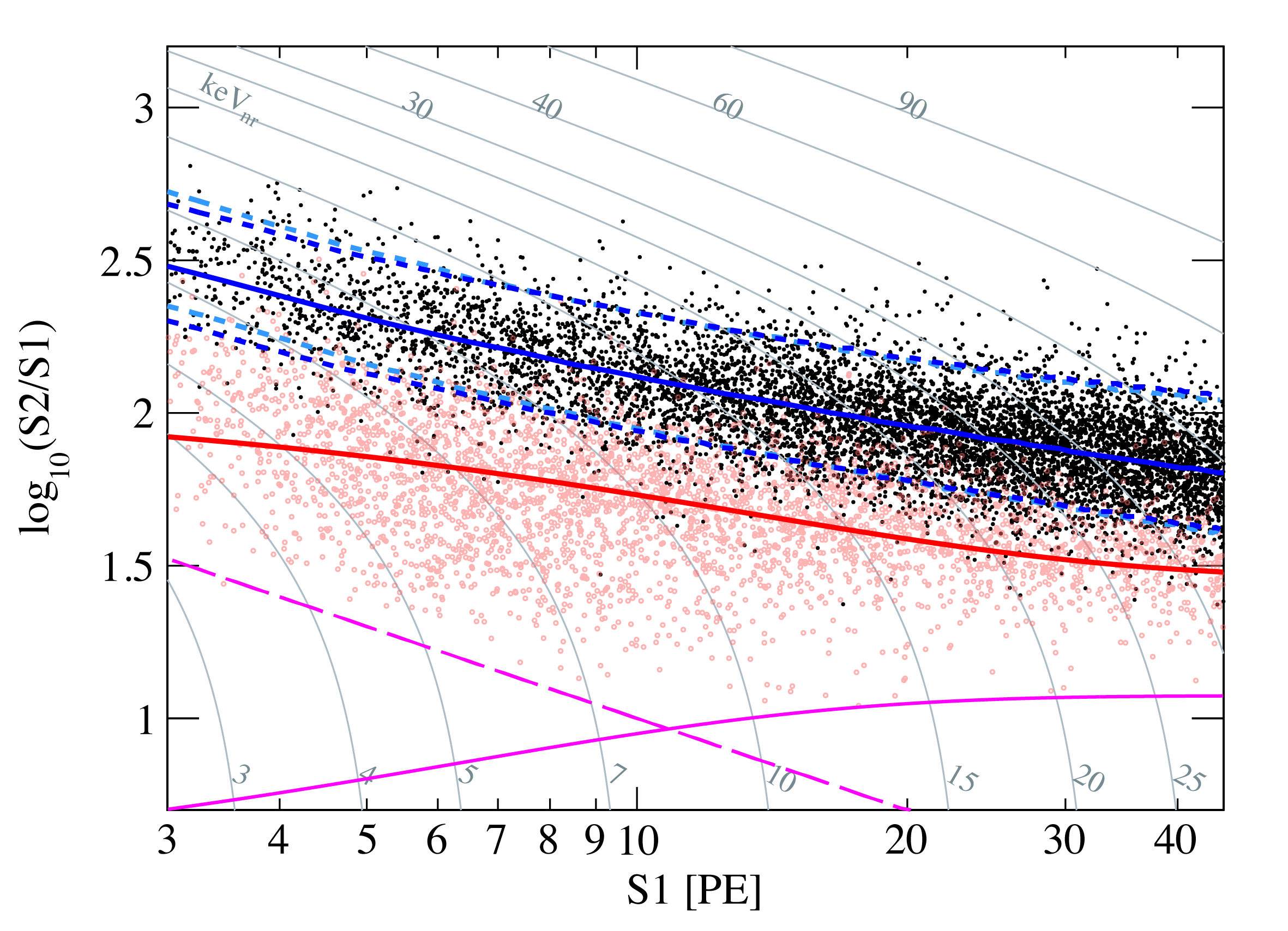}
  \caption{Tritium (solid black dots) and AmBe data (open red circles)
    in log$_{10}(S2/S1)$ vs. $S1$. For comparison, the median (Run 9
    and Run 10 averaged, solid blue), 10\% quantile, and 90\% quantile (Run 9:
    dashed blue; Run 10, light dashed blue) of the ER background PDFs
    are overlaid.  The solid red line is the median of the AmBe
    events. The dashed and solid magenta curves are the 100 PE
    selection cut for $S2$, and the 99.99\% NR acceptance curve from
    the MC calculation, respectively. The gray curves represent the
    equal energy curves in nuclear recoil energy (keV$_{\rm{nr}}$).}
  \label{fig:calibration}
\end{figure}

As described in Ref.~\cite{Tan:2016zwf}, to characterize the detector
response to ER background, about 1$\times10^{-15}$ mol of CH$_3$T was
loaded into the detector. The entire calibration run lasted for 44 days, but
only the data in the first 18 hours were used in the previous paper,
where the electron lifetime quickly deteriorated to an average
124~$\mu$s due to electronegative impurities. In this analysis, we
chose a later data set with an average electron lifetime of
706~$\mu$s, which contained about 7500 low energy ER events in the
fiducial volume (FV). The distribution of these events in
$\log_{10}(S2/S1)$ vs. $S1$ is shown in Fig.~\ref{fig:calibration},
with AmBe events overlaid. The NEST-based simulation mentioned above
was then compared to the data.  A tuning on $\alpha$ and a
recombination fluctuation parameter for ER was performed, to match the
median and width of the band for $S1>5$~PE. The ER detection
efficiency was then extracted using the method mentioned earlier, and
verified to be in agreement with that for NR. Below the reference NR
median line, 40 events were identified, corresponding to a leakage of
0.53$\pm0.22_{\rm stat+sys}$\%, in good agreement with the MC
prediction. The tuned NEST model was used to produce ER background
PDFs.

There was significant background reduction in Run 10. The $^{127}$Xe
which was present in Run 9 had mostly decayed away. Some new
$^{127}$Xe background was introduced from a fresh surface xenon
bottle, mixed in during the distillation. The average decay rate in
Run 10 was estimated based on the 408~keV ER peaks to be
32$\pm$6~$\mu$Bq/kg. The corresponding 5 keV ER background in the DM
search window is estimated to be 0.021 mDRU (1 mDRU = $10^{-3}$
evt/kg/day/keV) with 20\% fractional uncertainty. The Kr background
was measured in the data {\it{in-situ}} using the delayed
$\beta$-$\gamma$ coincidence. 13 events were found in the entire
580-kg sensitive volume.  Taking the coincidence selection efficiency
from the MC, and assuming a $2\times10^{-11}$ abundance of $^{85}$Kr,
the Kr concentration in Xe was 6.6$\pm$2.2 ppt, more than a factor of
six improvement from Run 9, contributing to an ER background 
0.20$\pm$0.07 mDRU. From the energy spectrum of single-scatter events,
a slight excess at low energy is consistent with a residual tritium
background of 0.27 mDRU (left float in later likelihood fit).  The Rn
background, estimated based on the $\beta$-$\alpha$ coincidence of
$^{214}$Bi-$^{214}$Po and $^{212}$Bi-$^{212}$Po, was 7.7~$\mu$Bq/kg
and 0.63~$\mu$Bq/kg, respectively, consistent with the values measured in
Run 9.  The levels of the ER background are summarized in
Table~\ref{tab:er_bkg_budget}.
\begin{table}[h]
  \begin{tabular}{ccc}
    \hline\hline
    Item & Run 9 (mDRU) & Run 10 (mDRU)\\\hline
    $^{85}$Kr & 1.19$\pm$0.20 & 0.20$\pm$0.07 \\
    $^{127}$Xe & 0.42$\pm$0.10 & 0.021$\pm$0.005 \\
    $^3$H & 0 & 0.27$\pm$0.08 \\
    $^{222}$Rn & 0.13$\pm$0.07 & 0.12$\pm$0.06  \\
    $^{220}$Rn & 0.01$\pm$0.01 & 0.02$\pm$0.01  \\
    ER (material) & 0.20$\pm$0.10 & 0.20$\pm$0.10\\
    Solar $\nu$ & 0.01 & 0.01 \\
    $^{136}$Xe & 0.0022 & 0.0022 \\
    \hline
    Total & 1.96$\pm$0.25 & 0.79$\pm$0.16\\\hline
  \end{tabular}
  \caption{Summary of ER backgrounds from different components in
    Run 9 and Run 10. The tritium background for Run 10 in the table is
    based on the best fit to the data.
  }
  \label{tab:er_bkg_budget}
\end{table}

The estimate of the accidental background has been improved in the
present analysis. A random trigger run was set up to estimate the
isolated $S1$ rate. The method in Run 9, searching for isolated
$S1$-like signals before single $S1$ events (no $S2$), was found to
be sometimes biased by real single scatter events whose $S2$s were
mis-identified as $S1$. Removing such effects reduced the isolated
$S1$ rate by 14\% to 1.6~Hz in Run 9. The isolated $S1$ rate in Run 10
was lowered to 0.4~Hz, possibly a direct consequence of the reduced
PMT gain and dark rate mentioned earlier. The same
boosted-decision-tree (BDT) cuts as in Run 9 (Ref.~\cite{Tan:2016zwf})
were used to suppress this background. The updated total
(below-NR-median) accidental background is 12.2 (0.8) and 3.5 (0.5)
for Run 9 and Run 10, respectively, with a 45\% uncertainty estimated
based on the variation of isolated $S1$ rate in a given run period.

The neutron background, dominated by the radioactivity of the PTFE
materials, was estimated following the same approach as in
Ref.~\cite{Tan:2016diz}, but with updated detector responses. The
uncertainty is estimated to be 50\% using the AmBe calibration data
based on the ratio of detected single scatter NR events to the 4.4 MeV
$\gamma$s.

\begin{figure}[!htbp]
  \centering
  \begin{subfigure}{1.0\linewidth}
    \includegraphics[width=0.95\linewidth]{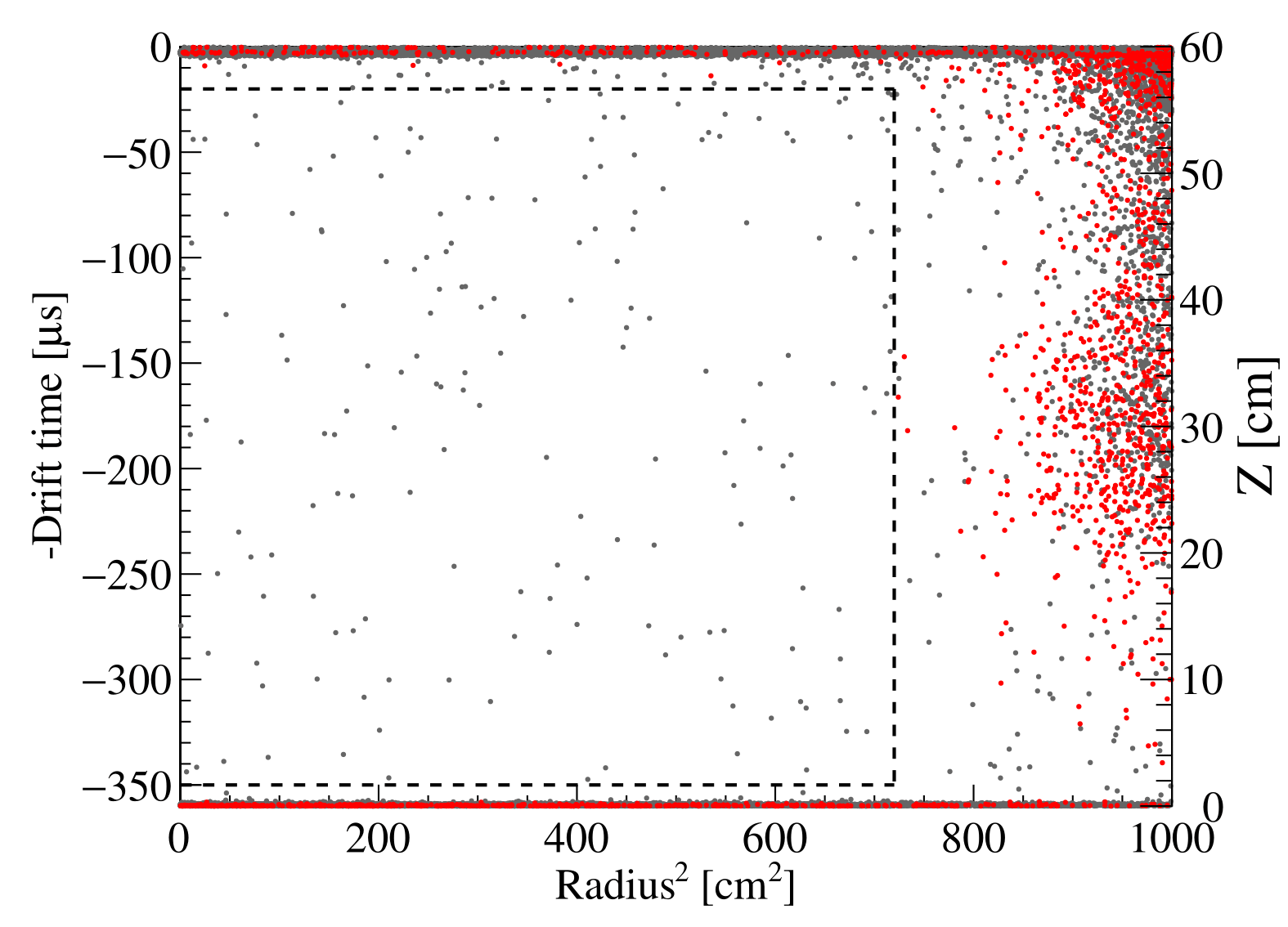}
    \caption{The $r^2$-$z$ distribution.}
    \label{fig:r2_z}
  \end{subfigure}
  \begin{subfigure}{1.0\linewidth}
    \includegraphics[width=0.95\linewidth]{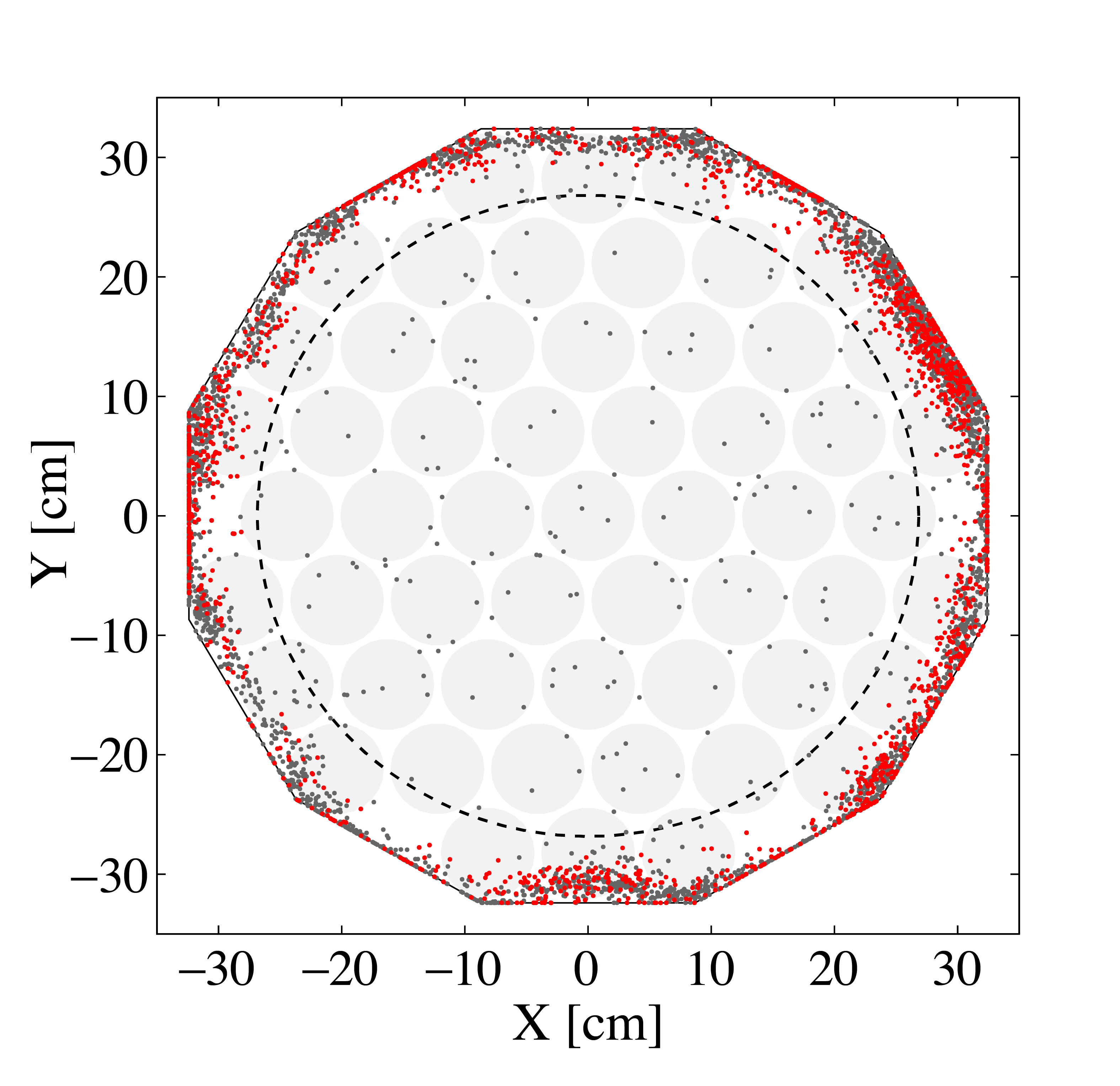}
    \caption{The $x$-$y$ distribution.}
    \label{fig:x_y}
  \end{subfigure}
  \caption{The position distributions for events within $S1$ and $S2$
    range cuts, (a) $z$ vs. $r^2$, and (b) $y$ vs. $x$. The drift time
    cut between 20 and 350~$\mu$s is applied for (b). The gray and red
    points are events above and below the NR median, respectively. The
    dashed box (a) and circle (b) represent the FV cut. Gray
    background circles in (b) indicate locations of the top PMTs.}
  \label{fig:final_candidate}
\end{figure}

The $S1$ and $S2$ range cuts were identical to those in Run 9,
i.e. from 3 to 45~PE for $S1$, 100 (raw) to 10000~PE for $S2$, and
above the 99.99\% NR acceptance curve in Fig.~\ref{fig:calibration}.
Remaining cuts were also kept the same to those in Run 9 except the
drift time cut. The lower cut was updated from 18 $\mu$s to 20 $\mu$s
by scaling with the new drift speed (weaker field). The upper cut of
350 $\mu$s (310 $\mu$s in Run 9) was chosen since the rate of
$^{127}$Xe-induced ``gamma-X'' events was reduced to be
negligible. The same radius-square cut $r^2<720$~cm$^2$ was used. The
FV was computed geometrically, and the drift field uniformity was
supported by the small position bias for events originated from the
wall. Within the FV, the target mass was 361.5$\pm$23.5~kg (328.9~kg
in Run 9) where the uncertainty was estimated using tritium and radon
events.  The survival events after successive cuts are shown in
Table~\ref{tab:eventrate}.
The vertex distribution of events falling into the $S1$ and $S2$
windows is shown in Fig.~\ref{fig:final_candidate}. The distribution
of events close to the PTFE wall with abnormally small $S2$ and
non-uniform vertical distribution were attributed to the loss of
electrons on the wall due to the local field irregularity.
In Fig.~\ref{fig:r2_z}, the red cluster close to $Z=30$~cm is due to
the peripheral top PMT with high noise and high ZLE threshold (located
in the upper right corner in Fig.~\ref{fig:x_y}), causing biased reconstructed
position particularly for wall events with suppressed $S2$ deep in the
TPC.
In the FV, the residual events are uniformly distributed. In total
there are 177 final candidate events.  The distribution of
log$_{10}(S2/S1)$ vs. $S1$ of these events is shown in
Fig.~\ref{fig:final_candidate}, mostly consistent with ER
background. For reference, no events are identified below the NR
median line, with 1.8$\pm$0.5 expected background, indicating a
downward fluctuation of background. In combination with the
below-NR-events in Run 9 (Table~\ref{tab:backgroundtable}), 
the probability of observing one or less events when 5 are expected is 7.2\%.

\begin{table}[h]
\centering
\begin{tabular}{ccc}
\hline\hline
Cut & run 9 & run 10 \\
\hline
All triggers & 24502402 & 18369083 \\
Quality cuts & 5160513 & 3070111 \\
$S1$ and $S2$ range  & 131097 & 111854 \\
FV cut & 398 & 178  \\
BDT cut & 389 & 177 \\
\hline\hline
\end{tabular}
\caption{Number of events in Run 9 and Run 10 after successive analysis selections.}
\label{tab:eventrate}
\end{table}

\begin{figure}[!htbp]
  \centering
  \includegraphics[width=0.95\linewidth]{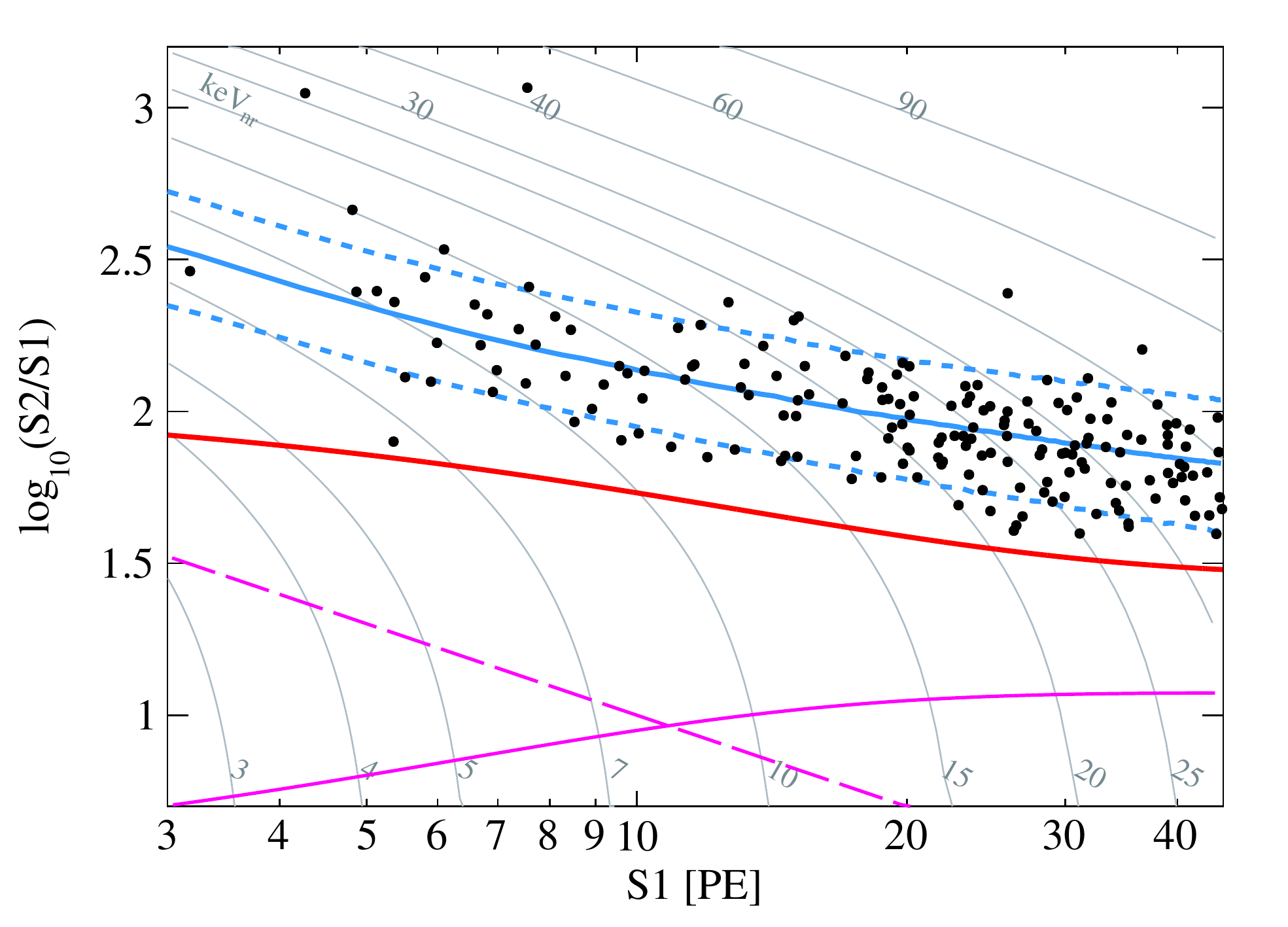}
  \caption{The distribution of log$_{10}$($S2$/$S1$) vs. $S1$ for the
    DM search data in Run 10, overlaid with the corresponding median,
    10\% quantile, and 90\% quantile of the ER background PDFs. The red curve
    is the NR median from AmBe calibration. }
  \label{fig:final_candidate_log}
\end{figure}

\begin{table}[h]
\centering
\begin{tabular}{cccc|c|c}
\hline\hline
 & ER & Accidental & Neutron & \parbox[t]{1.4cm}{Total\\Fitted} & \parbox[t]{1.4cm}{Total\\Observed}\\
\hline
Run 9 & 376.1 & 13.5 & 0.85 & 390$\pm$50 & 389 \\
\hline
\parbox{1.8cm}{Below \\NR median} & 2.0 & 0.9  & 0.35 & 3.2$\pm$0.9 & 1\\
\hline
Run 10 & 172.2 & 3.9 & 0.83 & 177$\pm$33 & 177 \\
\hline
\parbox{1.8cm}{Below \\NR median} & 0.9 & 0.6 & 0.33 & 1.8$\pm$0.5 & 0\\
\hline
\hline
\end{tabular}
\caption{The best fit total and below-NR-median background events in Run 9 and Run 10
  in the FV. The fractional uncertainties of expected events in the table
  are 13\% (Run 9 ER), 20\% (Run 10 ER),  45\% (accidental), and
  50\% (neutron), respectively, and propagated into that for the total fitted events.
  The below-NR-median ER background for Run 9 was updated using the new ER
  calibration. The corresponding best fit background
  nuisance parameters ($\delta_b$'s in Eqn.~\ref{eq:likelihood}) are
  0.123 ($^{127}$Xe), 0.135 (tritium), $-$0.105 (flat ER),
  0.111 (accidental), and $-$0.098 (neutron).
  Number of events from the data are shown in the
  last column.}
\label{tab:backgroundtable}
\end{table}

\begin{figure}[!htbp]
  \centering
  \begin{subfigure}{1.0\linewidth}
    \includegraphics[width=0.95\linewidth]{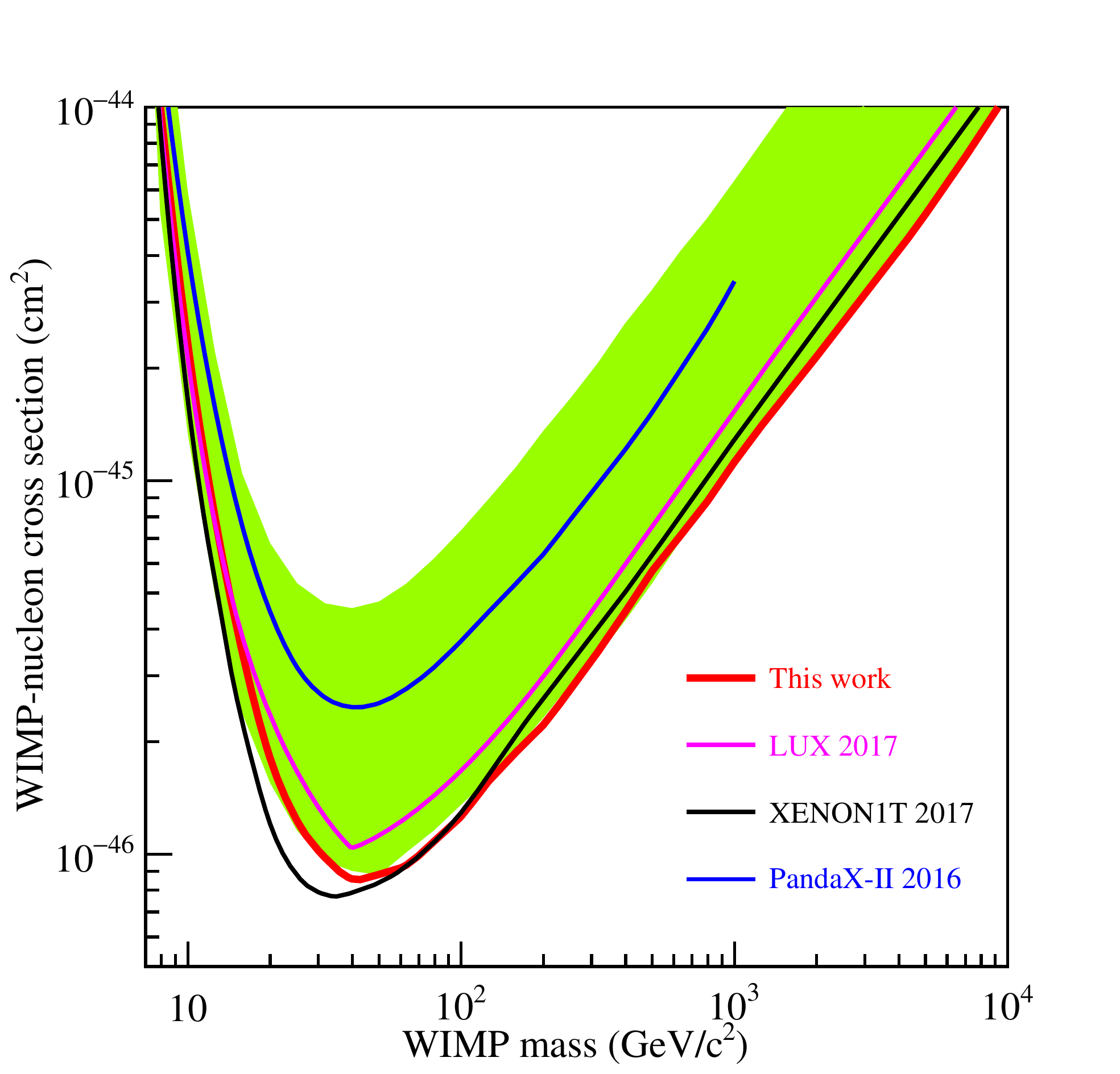}
    \caption{log scale in $m_{\chi}$}
    \label{fig:log_scale}
  \end{subfigure}
  \begin{subfigure}{1.0\linewidth}
    \includegraphics[width=0.95\linewidth]{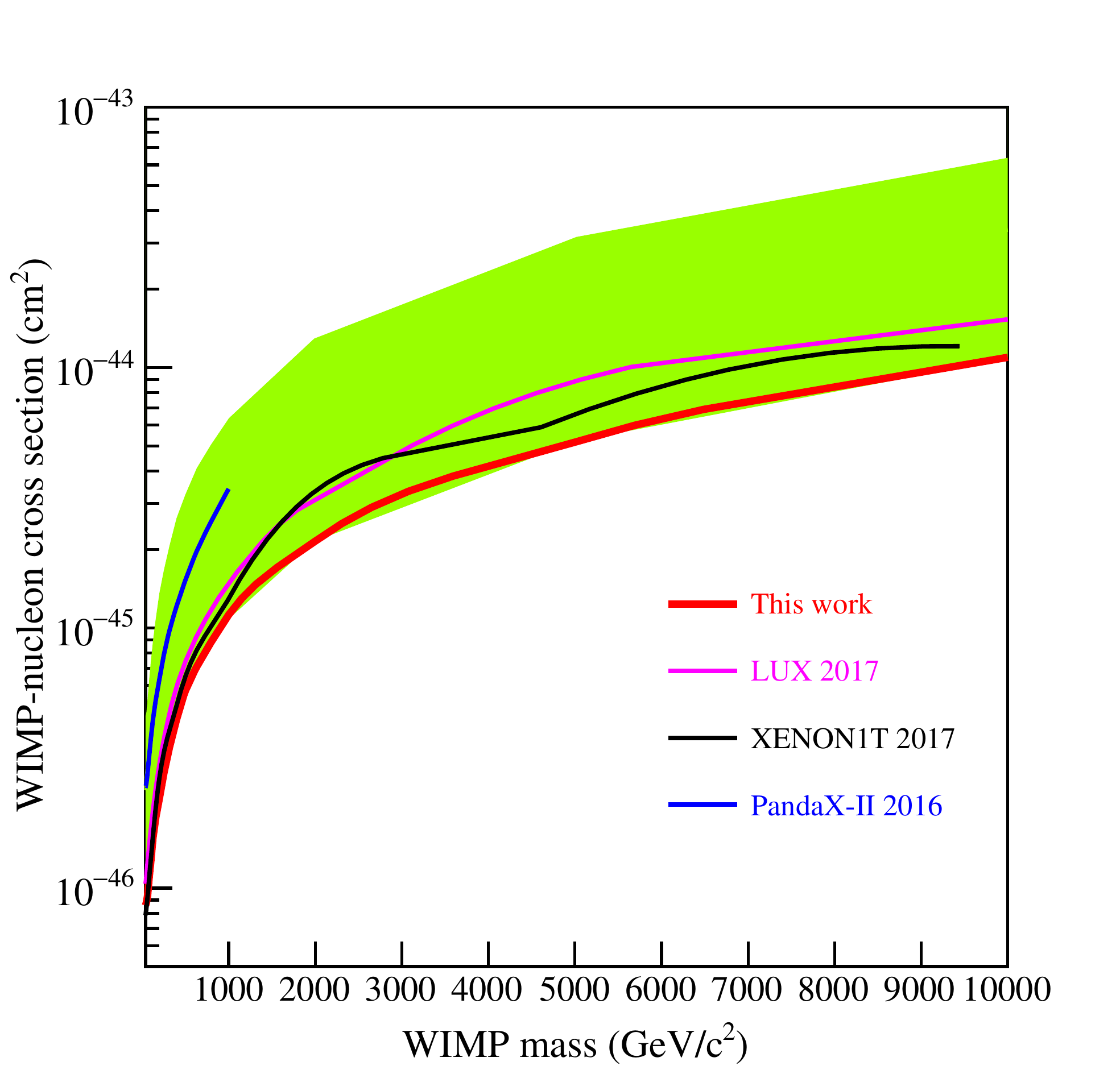}
    \caption{linear scale in $m_{\chi}$}
    \label{fig:lin_scale}
  \end{subfigure}
  \caption{The $90\%$ C.L. upper limits vs. $m_{\chi}$ [(a) log scale,
    (b) linear scale between 40 GeV/$c^2$ to 10 TeV/$c^2$] for the
    spin independent WIMP-nucleon elastic cross sections from the
    combined PandaX-II Run 9 and Run 10 data (red), overlaid with that
    from PandaX-II 2016~\cite{Tan:2016zwf} (blue), LUX
    2017~\cite{Akerib:2016vxi} (magenta), and XENON1T
    2017~\cite{Aprile:2017iyp} (black). The green band represents the
    $\pm1\sigma$ sensitivity band.}
  \label{fig:final_limit}
\end{figure}

The final candidates in Runs 9 and 10 were combined to search for
WIMPs.  An unbinned likelihood function was constructed as
\begin{equation}
\label{eq:likelihood}
  \mathcal{L}_{\rm pandax} = \big[\prod_{n=1}^{\textrm{nset}}\mathcal{L}_{n}\big] \times \big[G(\delta_{\rm DM}, \sigma_{\rm DM}) \prod_{b}G(\delta_b, \sigma_b)\big]\,,
\end{equation}
where
\begin{eqnarray}
  \mathcal{L}_{n} = &&{\rm Poiss}(N_{\rm meas}^n|N_{\rm fit}^n)\times \\\nonumber
  &&\Bigg[\displaystyle{\prod_{i=1}^{N_{\rm meas}^n}}\left(\frac{N_{\rm DM}^n(1+\delta_{\rm DM})P_{\rm DM}^n(S1^i,S2^i)}{N_{\rm fit}^n}\right. \\\nonumber
&&+ \left.\displaystyle\sum_b \frac{N_{b}^n(1+\delta_{b})P_{b}^n(S1^i,S2^i)}{N_{\rm fit}^n}\right)\Bigg]\,.
\end{eqnarray}
As in Ref.~\cite{Tan:2016zwf}, the data were divided into 14 sets in
Run 9, and 4 sets in Run 10 (nset = 18) to reflect different operation
conditions in TPC fields and electron lifetime. For each data set $n$,
$N_{\rm meas}^n$ and $N_{\rm fit}^n$ represent the measured and fitted
total numbers of detected events; $N_{\rm DM}^n$ and $N_{b}^n$ are the
numbers of WIMP and background events, with their corresponding PDFs
$P_{\rm DM}^n(S1,S2)$ and $P_{b}^n(S1,S2)$. The detection efficiencies
needed for determining the detected numbers of events are either
contained in the PDF, or included in $N_b$ (accidental
background). Five background components (represented by the subscript
``$b$'') are considered, including $^{127}$Xe, tritium, other flat ER
($^{85}$Kr, radon, and other detector gamma background), accidental,
and neutron background. Among all data sets, $\delta_b$ and $\sigma_b$
are the common nuisance normalization parameters and fractional
systematic uncertainties, respectively, with $\sigma_b$ taken from
Table~\ref{tab:er_bkg_budget}, and $G(\delta_b, \sigma_b)$ is the
Gaussian penalty term.  For WIMP detection, we also assumed an
normalization nuisance parameter $\delta_{\rm DM}$ constrained by a
$\sigma_{\rm DM}$ of 20\%, conservatively estimated using different NR
models as well as the uncertainties to PDE and EEE. The ER and NR
background PDFs were generated using the tuned NEST models, and the
accidental PDFs were produced from randomly paired data. The WIMP
spectrum was calculated using the same formalism as in
Ref.~\cite{Xiao:2014xyn, Xiao:2015psa}, including all the nuclear and
astronomical input parameters (standard isothermal halo model with a
DM density of 0.3 GeV/$c^2/\rm{cm}^3$). The WIMP PDFs at different masses
($m_{\chi}$) were produced using the tuned NR model with detection
efficiency embeded. For all WIMP masses between 5 GeV/$c^2$ to 10
TeV/$c^2$, the best fit cross section was always zero, and the best
fit nuisance parameters were all within $1\sigma$ from the nominal
values. The standard profile likelihood test statistic was used to set
the exclusion limit on the spin independent WIMP-nucleon elastic
scattering cross section ($\sigma_{\chi,n}$)~\cite{Cowan:2010js,
  Aprile:2011hx}.  The test statistic was calculated at grids of
($m_{\chi}$, $\sigma_{\chi,n}$) for the data, and compared to those
obtained from large number of toy Monte Carlo produced and fitted
using the same signal hypotheses~\cite{Feldman:1997qc}.
The final $90\%$ confidence level (C.L.) limit is shown in
Fig.~\ref{fig:final_limit}, together with limits from PandaX-II
2016~\cite{Tan:2016zwf}, LUX~\cite{Akerib:2016vxi}, and
XENON1T~\cite{Aprile:2017iyp} (see also Fig. 18 in Supplemental
Material [\cite{sup_material}]). The limit is very close to the
$-1\sigma$ of the sensitivity band for WIMP mass above 20 GeV/$c^2$,
therefore power constraining~\cite{Cowan:2011an} the limit to
$-1\sigma$ of the sensitivity would make little difference.
The strongest limit is set to be $8.6\times10^{-47}$ cm$^2$ at the
WIMP mass of 40 GeV/$c^2$. The limit curve corresponds to on average 2.3 signal
events across the full mass range, e.g. 1.9 at 10 GeV/$c^2$ and 2.6 at
1 TeV/$c^2$.
This limit is about a factor of three more constraining than our
previous results~\cite{Tan:2016zwf} (using the CL$_{s}$
approach~\cite{CLS1,CLS2}), and represents the most stringent limit on
elastic WIMP-nucleon spin-independent cross section for WIMP mass larger
than 100 GeV/$c^2$.

In summary, we report the combined WIMP search results using the data
with an exposure of 54 ton-day, the largest of its kind, from the
PandaX-II experiment.  Like the previous attempts, no WIMP candidates
have been identified. This yields a most stringent limit for
WIMP-nucleon cross section for masses larger than
100~GeV/$c^2$. Theoretical models indicate the importance of enhancing
the current search sensitivity by another order of magnitude.
PandaX-II detector will continue to run until a future upgrade to a
multi-ton scale experiment at CJPL.

\begin{acknowledgments}
  This project has been supported by a 985-III grant from Shanghai
  Jiao Tong University, grants from National Science Foundation of
  China (Nos. 11365022, 11435008, 11455001, 11505112 and 11525522), a
  grant from the Ministry of Science and Technology of China
  (No. 2016YFA0400301). We thank the support of grants from the Office
  of Science and Technology, Shanghai Municipal Government
  (No. 11DZ2260700, No. 16DZ2260200), and the support from the Key
  Laboratory for Particle Physics, Astrophysics and Cosmology,
  Ministry of Education. This work is supported in part by the Chinese
  Academy of Sciences Center for Excellence in Particle Physics
  (CCEPP) and Hongwen Foundation in Hong Kong.  We also would like to
  thank Dr. Xunhua Yuan and Chunfa Yao of China Iron and Steel
  Research Institute Group, and Taiyuan Iron and Steel (Group) Co. LTD
  for crucial help on low background stainless steel.  Finally, we
  thank the following organizations for indispensable logistics and
  other supports: the CJPL administration and the Yalong River
  Hydropower Development Company Ltd.
\end{acknowledgments}

\bibliographystyle{apsrev4-1}
\bibliography{refs.bib}

\begin{thebibliography}{25}%
\makeatletter
\providecommand \@ifxundefined [1]{%
 \@ifx{#1\undefined}
}%
\providecommand \@ifnum [1]{%
 \ifnum #1\expandafter \@firstoftwo
 \else \expandafter \@secondoftwo
 \fi
}%
\providecommand \@ifx [1]{%
 \ifx #1\expandafter \@firstoftwo
 \else \expandafter \@secondoftwo
 \fi
}%
\providecommand \natexlab [1]{#1}%
\providecommand \enquote  [1]{``#1''}%
\providecommand \bibnamefont  [1]{#1}%
\providecommand \bibfnamefont [1]{#1}%
\providecommand \citenamefont [1]{#1}%
\providecommand \href@noop [0]{\@secondoftwo}%
\providecommand \href [0]{\begingroup \@sanitize@url \@href}%
\providecommand \@href[1]{\@@startlink{#1}\@@href}%
\providecommand \@@href[1]{\endgroup#1\@@endlink}%
\providecommand \@sanitize@url [0]{\catcode `\\12\catcode `\$12\catcode
  `\&12\catcode `\#12\catcode `\^12\catcode `\_12\catcode `\%12\relax}%
\providecommand \@@startlink[1]{}%
\providecommand \@@endlink[0]{}%
\providecommand \url  [0]{\begingroup\@sanitize@url \@url }%
\providecommand \@url [1]{\endgroup\@href {#1}{\urlprefix }}%
\providecommand \urlprefix  [0]{URL }%
\providecommand \Eprint [0]{\href }%
\providecommand \doibase [0]{http://dx.doi.org/}%
\providecommand \selectlanguage [0]{\@gobble}%
\providecommand \bibinfo  [0]{\@secondoftwo}%
\providecommand \bibfield  [0]{\@secondoftwo}%
\providecommand \translation [1]{[#1]}%
\providecommand \BibitemOpen [0]{}%
\providecommand \bibitemStop [0]{}%
\providecommand \bibitemNoStop [0]{.\EOS\space}%
\providecommand \EOS [0]{\spacefactor3000\relax}%
\providecommand \BibitemShut  [1]{\csname bibitem#1\endcsname}%
\let\auto@bib@innerbib\@empty
\bibitem [{\citenamefont {Bertone}\ \emph {et~al.}(2005)\citenamefont
  {Bertone}, \citenamefont {Hooper},\ and\ \citenamefont
  {Silk}}]{Bertone:2004pz}%
  \BibitemOpen
  \bibfield  {author} {\bibinfo {author} {\bibfnamefont {G.}~\bibnamefont
  {Bertone}}, \bibinfo {author} {\bibfnamefont {D.}~\bibnamefont {Hooper}}, \
  and\ \bibinfo {author} {\bibfnamefont {J.}~\bibnamefont {Silk}},\ }\href
  {\doibase 10.1016/j.physrep.2004.08.031} {\bibfield  {journal} {\bibinfo
  {journal} {Phys. Rept.}\ }\textbf {\bibinfo {volume} {405}},\ \bibinfo
  {pages} {279} (\bibinfo {year} {2005})},\ \Eprint
  {http://arxiv.org/abs/hep-ph/0404175} {arXiv:hep-ph/0404175 [hep-ph]}
  \BibitemShut {NoStop}%
\bibitem [{\citenamefont {Akerib}\ \emph {et~al.}(2017)\citenamefont {Akerib}
  \emph {et~al.}}]{Akerib:2016vxi}%
  \BibitemOpen
  \bibfield  {author} {\bibinfo {author} {\bibfnamefont {D.~S.}\ \bibnamefont
  {Akerib}} \emph {et~al.} (\bibinfo {collaboration} {LUX}),\ }\href {\doibase
  10.1103/PhysRevLett.118.021303} {\bibfield  {journal} {\bibinfo  {journal}
  {Phys. Rev. Lett.}\ }\textbf {\bibinfo {volume} {118}},\ \bibinfo {pages}
  {021303} (\bibinfo {year} {2017})},\ \Eprint
  {http://arxiv.org/abs/1608.07648} {arXiv:1608.07648 [astro-ph.CO]}
  \BibitemShut {NoStop}%
\bibitem [{\citenamefont {Tan}\ \emph {et~al.}(2016{\natexlab{a}})\citenamefont
  {Tan} \emph {et~al.}}]{Tan:2016zwf}%
  \BibitemOpen
  \bibfield  {author} {\bibinfo {author} {\bibfnamefont {A.}~\bibnamefont
  {Tan}} \emph {et~al.} (\bibinfo {collaboration} {PandaX-II}),\ }\href
  {\doibase 10.1103/PhysRevLett.117.121303} {\bibfield  {journal} {\bibinfo
  {journal} {Phys. Rev. Lett.}\ }\textbf {\bibinfo {volume} {117}},\ \bibinfo
  {pages} {121303} (\bibinfo {year} {2016}{\natexlab{a}})},\ \Eprint
  {http://arxiv.org/abs/1607.07400} {arXiv:1607.07400 [hep-ex]} \BibitemShut
  {NoStop}%
\bibitem [{\citenamefont {Aprile}\ \emph {et~al.}(2017)\citenamefont {Aprile}
  \emph {et~al.}}]{Aprile:2017iyp}%
  \BibitemOpen
  \bibfield  {author} {\bibinfo {author} {\bibfnamefont {E.}~\bibnamefont
  {Aprile}} \emph {et~al.} (\bibinfo {collaboration} {XENON}),\ }\href@noop {}
  {\  (\bibinfo {year} {2017})},\ \Eprint {http://arxiv.org/abs/1705.06655}
  {arXiv:1705.06655 [astro-ph.CO]} \BibitemShut {NoStop}%
\bibitem [{\citenamefont {Bagnaschi}\ \emph {et~al.}(2015)\citenamefont
  {Bagnaschi} \emph {et~al.}}]{Bagnaschi:2015eha}%
  \BibitemOpen
  \bibfield  {author} {\bibinfo {author} {\bibfnamefont {E.~A.}\ \bibnamefont
  {Bagnaschi}} \emph {et~al.},\ }\href {\doibase
  10.1140/epjc/s10052-015-3718-9} {\bibfield  {journal} {\bibinfo  {journal}
  {Eur. Phys. J.}\ }\textbf {\bibinfo {volume} {C75}},\ \bibinfo {pages} {500}
  (\bibinfo {year} {2015})},\ \Eprint {http://arxiv.org/abs/1508.01173}
  {arXiv:1508.01173 [hep-ph]} \BibitemShut {NoStop}%
\bibitem [{\citenamefont {Liu}\ \emph {et~al.}(2017)\citenamefont {Liu},
  \citenamefont {Chen},\ and\ \citenamefont {Ji}}]{Liu:2017drf}%
  \BibitemOpen
  \bibfield  {author} {\bibinfo {author} {\bibfnamefont {J.}~\bibnamefont
  {Liu}}, \bibinfo {author} {\bibfnamefont {X.}~\bibnamefont {Chen}}, \ and\
  \bibinfo {author} {\bibfnamefont {X.}~\bibnamefont {Ji}},\ }\href {\doibase
  10.1038/nphys4039} {\bibfield  {journal} {\bibinfo  {journal} {Nature Phys.}\
  }\textbf {\bibinfo {volume} {13}},\ \bibinfo {pages} {212} (\bibinfo {year}
  {2017})}\BibitemShut {NoStop}%
\bibitem [{\citenamefont {Kang}\ \emph {et~al.}(2010)\citenamefont {Kang},
  \citenamefont {Cheng}, \citenamefont {Chen}, \citenamefont {Li},
  \citenamefont {Shen}, \citenamefont {Wu},\ and\ \citenamefont
  {Yue}}]{Kang:2010zza}%
  \BibitemOpen
  \bibfield  {author} {\bibinfo {author} {\bibfnamefont {K.~J.}\ \bibnamefont
  {Kang}}, \bibinfo {author} {\bibfnamefont {J.~P.}\ \bibnamefont {Cheng}},
  \bibinfo {author} {\bibfnamefont {Y.~H.}\ \bibnamefont {Chen}}, \bibinfo
  {author} {\bibfnamefont {Y.~J.}\ \bibnamefont {Li}}, \bibinfo {author}
  {\bibfnamefont {M.~B.}\ \bibnamefont {Shen}}, \bibinfo {author}
  {\bibfnamefont {S.~Y.}\ \bibnamefont {Wu}}, \ and\ \bibinfo {author}
  {\bibfnamefont {Q.}~\bibnamefont {Yue}},\ }\bibfield  {booktitle} {\emph
  {\bibinfo {booktitle} {{Proceedings, 11th International Conference on Topics
  in astroparticle and underground physics in Memory of Julio Morales (TAUP
  2009): Rome, Italy, July 1-5, 2009}}},\ }\href {\doibase
  10.1088/1742-6596/203/1/012028} {\bibfield  {journal} {\bibinfo  {journal}
  {J. Phys. Conf. Ser.}\ }\textbf {\bibinfo {volume} {203}},\ \bibinfo {pages}
  {012028} (\bibinfo {year} {2010})}\BibitemShut {NoStop}%
\bibitem [{\citenamefont {Tan}\ \emph {et~al.}(2016{\natexlab{b}})\citenamefont
  {Tan} \emph {et~al.}}]{Tan:2016diz}%
  \BibitemOpen
  \bibfield  {author} {\bibinfo {author} {\bibfnamefont {A.}~\bibnamefont
  {Tan}} \emph {et~al.} (\bibinfo {collaboration} {PandaX}),\ }\href {\doibase
  10.1103/PhysRevD.93.122009} {\bibfield  {journal} {\bibinfo  {journal} {Phys.
  Rev.}\ }\textbf {\bibinfo {volume} {D93}},\ \bibinfo {pages} {122009}
  (\bibinfo {year} {2016}{\natexlab{b}})},\ \Eprint
  {http://arxiv.org/abs/1602.06563} {arXiv:1602.06563 [hep-ex]} \BibitemShut
  {NoStop}%
\bibitem [{\citenamefont {Akerib}\ \emph {et~al.}(2016)\citenamefont {Akerib}
  \emph {et~al.}}]{Akerib:2015wdi}%
  \BibitemOpen
  \bibfield  {author} {\bibinfo {author} {\bibfnamefont {D.~S.}\ \bibnamefont
  {Akerib}} \emph {et~al.} (\bibinfo {collaboration} {LUX}),\ }\href {\doibase
  10.1103/PhysRevD.93.072009} {\bibfield  {journal} {\bibinfo  {journal} {Phys.
  Rev.}\ }\textbf {\bibinfo {volume} {D93}},\ \bibinfo {pages} {072009}
  (\bibinfo {year} {2016})},\ \Eprint {http://arxiv.org/abs/1512.03133}
  {arXiv:1512.03133 [physics.ins-det]} \BibitemShut {NoStop}%
\bibitem [{v17()}]{v1724_reference}%
  \BibitemOpen
  \href@noop {} {}\bibinfo {note}
  {\url{http://www.caen.it/servlet/checkCaenManualFile?Id=12364}}\BibitemShut
  {NoStop}%
\bibitem [{sup()}]{sup_material}%
  \BibitemOpen
  \href@noop {} {}\bibinfo {note} {Supplemental Material, see
  \url{https://pandax.sjtu.edu.cn/articles/2nd/supplemental.pdf}, which
  includes Refs.~\cite{Akerib:2016vxi}, \cite{Tan:2016zwf},
  \cite{Aprile:2017iyp} and \cite{Ji:2017tevpa}}\BibitemShut {NoStop}%
\bibitem [{\citenamefont {Wu}\ \emph {et~al.}(2017)\citenamefont {Wu} \emph
  {et~al.}}]{Wu:2017cjl}%
  \BibitemOpen
  \bibfield  {author} {\bibinfo {author} {\bibfnamefont {Q.}~\bibnamefont {Wu}}
  \emph {et~al.},\ }\href {\doibase 10.1088/1748-0221/12/08/T08004} {\bibfield
  {journal} {\bibinfo  {journal} {JINST}\ }\textbf {\bibinfo {volume} {12}},\
  \bibinfo {pages} {T08004} (\bibinfo {year} {2017})},\ \Eprint
  {http://arxiv.org/abs/1707.02134} {arXiv:1707.02134 [physics.ins-det]}
  \BibitemShut {NoStop}%
\bibitem [{\citenamefont {Agostinelli}\ \emph {et~al.}(2003)\citenamefont
  {Agostinelli} \emph {et~al.}}]{Agostinelli:2002hh}%
  \BibitemOpen
  \bibfield  {author} {\bibinfo {author} {\bibfnamefont {S.}~\bibnamefont
  {Agostinelli}} \emph {et~al.} (\bibinfo {collaboration} {GEANT4}),\ }\href
  {\doibase 10.1016/S0168-9002(03)01368-8} {\bibfield  {journal} {\bibinfo
  {journal} {Nucl. Instrum. Meth.}\ }\textbf {\bibinfo {volume} {A506}},\
  \bibinfo {pages} {250} (\bibinfo {year} {2003})}\BibitemShut {NoStop}%
\bibitem [{\citenamefont {Allison}\ \emph {et~al.}(2006)\citenamefont {Allison}
  \emph {et~al.}}]{Allison:2006ve}%
  \BibitemOpen
  \bibfield  {author} {\bibinfo {author} {\bibfnamefont {J.}~\bibnamefont
  {Allison}} \emph {et~al.},\ }\href {\doibase 10.1109/TNS.2006.869826}
  {\bibfield  {journal} {\bibinfo  {journal} {IEEE Trans. Nucl. Sci.}\ }\textbf
  {\bibinfo {volume} {53}},\ \bibinfo {pages} {270} (\bibinfo {year}
  {2006})}\BibitemShut {NoStop}%
\bibitem [{\citenamefont {Lenardo}\ \emph {et~al.}(2015)\citenamefont
  {Lenardo}, \citenamefont {Kazkaz}, \citenamefont {Szydagis},\ and\
  \citenamefont {Tripathi}}]{Lenardo:2014cva}%
  \BibitemOpen
  \bibfield  {author} {\bibinfo {author} {\bibfnamefont {B.}~\bibnamefont
  {Lenardo}}, \bibinfo {author} {\bibfnamefont {K.}~\bibnamefont {Kazkaz}},
  \bibinfo {author} {\bibfnamefont {M.}~\bibnamefont {Szydagis}}, \ and\
  \bibinfo {author} {\bibfnamefont {M.}~\bibnamefont {Tripathi}},\ }\href
  {\doibase 10.1109/TNS.2015.2481322} {\bibfield  {journal} {\bibinfo
  {journal} {IEEE Trans. Nucl. Sci.}\ }\textbf {\bibinfo {volume} {62}},\
  \bibinfo {pages} {3387} (\bibinfo {year} {2015})},\ \Eprint
  {http://arxiv.org/abs/1412.4417} {arXiv:1412.4417 [astro-ph.IM]} \BibitemShut
  {NoStop}%
\bibitem [{\citenamefont {Faham}\ \emph {et~al.}(2015)\citenamefont {Faham},
  \citenamefont {Gehman}, \citenamefont {Currie}, \citenamefont {Dobi},
  \citenamefont {Sorensen},\ and\ \citenamefont {Gaitskell}}]{Faham:2015kqa}%
  \BibitemOpen
  \bibfield  {author} {\bibinfo {author} {\bibfnamefont {C.~H.}\ \bibnamefont
  {Faham}}, \bibinfo {author} {\bibfnamefont {V.~M.}\ \bibnamefont {Gehman}},
  \bibinfo {author} {\bibfnamefont {A.}~\bibnamefont {Currie}}, \bibinfo
  {author} {\bibfnamefont {A.}~\bibnamefont {Dobi}}, \bibinfo {author}
  {\bibfnamefont {P.}~\bibnamefont {Sorensen}}, \ and\ \bibinfo {author}
  {\bibfnamefont {R.~J.}\ \bibnamefont {Gaitskell}},\ }\href {\doibase
  10.1088/1748-0221/10/09/P09010, 10.1088/1748-0221/2015/9/P09010} {\bibfield
  {journal} {\bibinfo  {journal} {JINST}\ }\textbf {\bibinfo {volume} {10}},\
  \bibinfo {pages} {P09010} (\bibinfo {year} {2015})},\ \Eprint
  {http://arxiv.org/abs/1506.08748} {arXiv:1506.08748 [physics.ins-det]}
  \BibitemShut {NoStop}%
\bibitem [{\citenamefont {Xiao}\ \emph {et~al.}(2014)\citenamefont {Xiao} \emph
  {et~al.}}]{Xiao:2014xyn}%
  \BibitemOpen
  \bibfield  {author} {\bibinfo {author} {\bibfnamefont {M.}~\bibnamefont
  {Xiao}} \emph {et~al.} (\bibinfo {collaboration} {PandaX}),\ }\href {\doibase
  10.1007/s11433-014-5598-7} {\bibfield  {journal} {\bibinfo  {journal} {Sci.
  China Phys. Mech. Astron.}\ }\textbf {\bibinfo {volume} {57}},\ \bibinfo
  {pages} {2024} (\bibinfo {year} {2014})},\ \Eprint
  {http://arxiv.org/abs/1408.5114} {arXiv:1408.5114 [hep-ex]} \BibitemShut
  {NoStop}%
\bibitem [{\citenamefont {Xiao}\ \emph {et~al.}(2015)\citenamefont {Xiao} \emph
  {et~al.}}]{Xiao:2015psa}%
  \BibitemOpen
  \bibfield  {author} {\bibinfo {author} {\bibfnamefont {X.}~\bibnamefont
  {Xiao}} \emph {et~al.} (\bibinfo {collaboration} {PandaX}),\ }\href {\doibase
  10.1103/PhysRevD.92.052004} {\bibfield  {journal} {\bibinfo  {journal} {Phys.
  Rev.}\ }\textbf {\bibinfo {volume} {D92}},\ \bibinfo {pages} {052004}
  (\bibinfo {year} {2015})},\ \Eprint {http://arxiv.org/abs/1505.00771}
  {arXiv:1505.00771 [hep-ex]} \BibitemShut {NoStop}%
\bibitem [{\citenamefont {Cowan}\ \emph
  {et~al.}(2011{\natexlab{a}})\citenamefont {Cowan}, \citenamefont {Cranmer},
  \citenamefont {Gross},\ and\ \citenamefont {Vitells}}]{Cowan:2010js}%
  \BibitemOpen
  \bibfield  {author} {\bibinfo {author} {\bibfnamefont {G.}~\bibnamefont
  {Cowan}}, \bibinfo {author} {\bibfnamefont {K.}~\bibnamefont {Cranmer}},
  \bibinfo {author} {\bibfnamefont {E.}~\bibnamefont {Gross}}, \ and\ \bibinfo
  {author} {\bibfnamefont {O.}~\bibnamefont {Vitells}},\ }\href {\doibase
  10.1140/epjc/s10052-011-1554-0, 10.1140/epjc/s10052-013-2501-z} {\bibfield
  {journal} {\bibinfo  {journal} {Eur. Phys. J.}\ }\textbf {\bibinfo {volume}
  {C71}},\ \bibinfo {pages} {1554} (\bibinfo {year} {2011}{\natexlab{a}})},\
  \bibinfo {note} {[Erratum: Eur. Phys. J.C73,2501(2013)]},\ \Eprint
  {http://arxiv.org/abs/1007.1727} {arXiv:1007.1727 [physics.data-an]}
  \BibitemShut {NoStop}%
\bibitem [{\citenamefont {Aprile}\ \emph {et~al.}(2011)\citenamefont {Aprile}
  \emph {et~al.}}]{Aprile:2011hx}%
  \BibitemOpen
  \bibfield  {author} {\bibinfo {author} {\bibfnamefont {E.}~\bibnamefont
  {Aprile}} \emph {et~al.} (\bibinfo {collaboration} {XENON100}),\ }\href
  {\doibase 10.1103/PhysRevD.84.052003} {\bibfield  {journal} {\bibinfo
  {journal} {Phys. Rev.}\ }\textbf {\bibinfo {volume} {D84}},\ \bibinfo {pages}
  {052003} (\bibinfo {year} {2011})},\ \Eprint {http://arxiv.org/abs/1103.0303}
  {arXiv:1103.0303 [hep-ex]} \BibitemShut {NoStop}%
\bibitem [{\citenamefont {Feldman}\ and\ \citenamefont
  {Cousins}(1998)}]{Feldman:1997qc}%
  \BibitemOpen
  \bibfield  {author} {\bibinfo {author} {\bibfnamefont {G.~J.}\ \bibnamefont
  {Feldman}}\ and\ \bibinfo {author} {\bibfnamefont {R.~D.}\ \bibnamefont
  {Cousins}},\ }\href {\doibase 10.1103/PhysRevD.57.3873} {\bibfield  {journal}
  {\bibinfo  {journal} {Phys. Rev.}\ }\textbf {\bibinfo {volume} {D57}},\
  \bibinfo {pages} {3873} (\bibinfo {year} {1998})},\ \Eprint
  {http://arxiv.org/abs/physics/9711021} {arXiv:physics/9711021
  [physics.data-an]} \BibitemShut {NoStop}%
\bibitem [{\citenamefont {Cowan}\ \emph
  {et~al.}(2011{\natexlab{b}})\citenamefont {Cowan}, \citenamefont {Cranmer},
  \citenamefont {Gross},\ and\ \citenamefont {Vitells}}]{Cowan:2011an}%
  \BibitemOpen
  \bibfield  {author} {\bibinfo {author} {\bibfnamefont {G.}~\bibnamefont
  {Cowan}}, \bibinfo {author} {\bibfnamefont {K.}~\bibnamefont {Cranmer}},
  \bibinfo {author} {\bibfnamefont {E.}~\bibnamefont {Gross}}, \ and\ \bibinfo
  {author} {\bibfnamefont {O.}~\bibnamefont {Vitells}},\ }\href@noop {} {\
  (\bibinfo {year} {2011}{\natexlab{b}})},\ \Eprint
  {http://arxiv.org/abs/1105.3166} {arXiv:1105.3166 [physics.data-an]}
  \BibitemShut {NoStop}%
\bibitem [{\citenamefont {Read}(2002)}]{CLS1}%
  \BibitemOpen
  \bibfield  {author} {\bibinfo {author} {\bibfnamefont {A.~L.}\ \bibnamefont
  {Read}},\ }\href {\doibase 10.1088/0954-3899/28/10/313} {\bibfield  {journal}
  {\bibinfo  {journal} {J. Phys.}\ }\textbf {\bibinfo {volume} {G28}},\
  \bibinfo {pages} {2693} (\bibinfo {year} {2002})}\BibitemShut {NoStop}%
\bibitem [{\citenamefont {Junk}(1999)}]{CLS2}%
  \BibitemOpen
  \bibfield  {author} {\bibinfo {author} {\bibfnamefont {T.}~\bibnamefont
  {Junk}},\ }\href@noop {} {\bibfield  {journal} {\bibinfo  {journal} {Nucl.
  Instrum. Meth.}\ }\textbf {\bibinfo {volume} {A434}},\ \bibinfo {pages} {435}
  (\bibinfo {year} {1999})},\ \Eprint {http://arxiv.org/abs/9902006}
  {arXiv:9902006 [hep-ex]} \BibitemShut {NoStop}%
\bibitem [{\citenamefont {Ji}(2017)}]{Ji:2017tevpa}%
  \BibitemOpen
  \bibfield  {author} {\bibinfo {author} {\bibfnamefont {X.}~\bibnamefont
  {Ji}},\ }in\ \href@noop {} {\emph {\bibinfo {booktitle} {TeV Particle
  Astrophysics 2017}}}\ (\bibinfo {address} {Columbus, Ohio, US},\ \bibinfo
  {year} {2017})\ \bibinfo {note}
  {\url{https://tevpa2017.osu.edu/talks/ji.pdf}}\BibitemShut {NoStop}%
\end{thebibliography}%

\end{document}